\documentclass[aps,prl,twocolumn,showpacs,superscriptaddress]{revtex4-2}
\usepackage{amsmath,amssymb}
\usepackage{graphicx}
\usepackage{wasysym}
\usepackage{amsfonts}
\usepackage{bm}
\usepackage{enumerate}
\usepackage{color}
\usepackage{url}

\usepackage{hyperref}
\hypersetup{
	colorlinks = true,
	urlcolor = blue,
	linkcolor = blue,
	citecolor = blue
}

\usepackage{epstopdf}
\usepackage{latexsym}
\usepackage[caption=false,singlelinecheck=false]{subfig}

\newcommand{\ket}[1]{\ensuremath{| #1 \rangle}}

\newcommand{\bea}{\begin{eqnarray}}
	\newcommand{\eea}{\end{eqnarray}}
\newcommand{\be}{\begin{eqnarray}}
	\newcommand{\ee}{\end{eqnarray}}
\newcommand{\bw}{\begin{widetext}}
	\newcommand{\ew}{\end{widetext}}

\newcommand{\bs}{\boldsymbol}

\begin{document}

\title{Kekule spin-orbit dimer phase and triplon dynamics}
	
\author{GiBaik Sim}
\affiliation {Department of Physics TQM, Technische Universit\"{a}t M\"{u}nchen, $\&$ James-Franck-Straße 1, D-85748 Garching, Germany}
\affiliation{Munich Center for Quantum Science and Technology (MCQST), 80799 Munich, Germany}
	
\begin{abstract}
	We derive and study a spin-orbital model for ions with $d^1$ electronic configuration on a honeycomb lattice. In this system, the directional character of $t_{2g}$ orbital leads to extensively degenerate dimerized ground states. We find that additional interactions from charge transfer processes completely lift the degeneracy and stabilize the Kekule spin-orbit dimerized phase where dimers form a kagome superlattice. For such phase, the triplon band spectrum resembles the electronic band structure of the kagome lattice and becomes topologically non-trivial in the presence of inter-dimer Dzyaloshinskii-Moriya interactions. As an experimental verification of the Kekule dimerized phase, we propose the thermal Hall experiment, which can directly uncover the topological profile of the corresponding triplon band spectrum.
\end{abstract}
	
\date{\today}
	
\maketitle
	
Isotropic spin models on two (three) dimensional lattice often have long-range ordered ground states with gapless collective excitation modes \cite{auerbach2012interacting}. However, this scenario fails for the geometrically frustrated cases whose classical ground states are (sub-)extensively degenerate \cite{henley1989ordering,baskaran2008spin,rousochatzakis2017classical}. For such cases, the ground state manifold responds to the following effects in various ways: It can become a highly entangled quantum spin liquid state with quantum fluctuations \cite{canals1998pyrochlore, baskaran2007exact}, lift the degeneracy through the order by disorder mechanism \cite{bergman2007order, lee2014order}, which often selects less entangled magnetically ordered state, or encounter structural phase transition \cite{yamashita2000spin, tchernyshyov2002order}, which reduces the symmetry of the lattice. 
	
The concept of classical degeneracy can be applied to a spin-orbital model where the ground state manifold is formed by an extensive number of dimer coverings \cite{jackeli2007dimer}. Here, the degeneracy originates from the directional character of $t_{2g}$ orbitals, which frustrate the spin interactions even on bipartite lattices. Recently, the spin-orbital model has been applied to many materials with dimerized ground state such as $\alpha-$RuCl$_3$ in the presence of high pressure \cite{bastien2018pressure}, Ba$_2$YMoO$_6$ with magnetically active Mo$^{5+}$ ($d^1$) ions on fcc lattice \cite{de2010valence,romhanyi2017spin}, and Li$_2$RuO$_3$ with Ru$^{4+}$ ($d^2$) ions on honeycomb lattice \cite{miura2007new}. In fact, Li$_2$RuO$_3$ shows a specific crystallized dimer ground state with a herringbone pattern which is well captured by the presence of magnetoelastic coupling \cite{jackeli2008classical}.
	
In this work, we point out an electronic degeneracy lifting mechanism that can be relevant to $\alpha-$MX$_3$ with M = Ti, Zr, Hf and X = F, Cl, Br where cation ions (M) with $d^1$ electronic configuration occupy a honeycomb lattice \cite{b1964synthesis}. By introducing charge transfer contributions, we show that the degeneracy of dimer covering is completely lifted and the system stabilizes Kekule spin-orbit dimerized phase. To investigate the dynamics of the phase, we introduce the triplon operator and investigate the corresponding triplon band spectrum, which turns out to be topologically non-trivial even with infinitesimal Dzyaloshinskii-Moriya interactions (DMI). We also investigate the thermal response and find that the topological properties are directly encoded in the thermal Hall conductivity.
	
\begin{figure}[t!]
	\includegraphics[scale=1.]{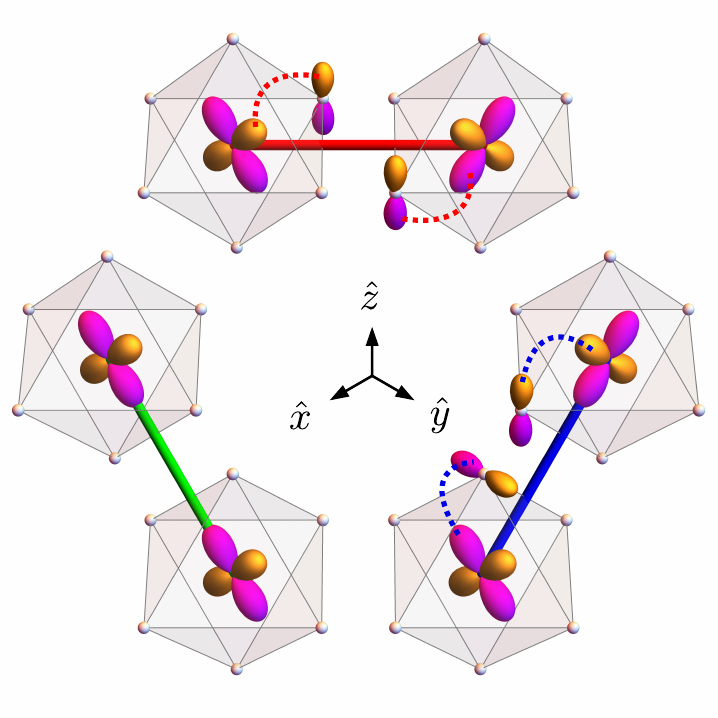}
	\caption{(color online) Possible hopping processes in $\alpha-$MX$_3$ which includes direct hopping (green), charge transfer (blue), and cyclic exchange (red) contribution. The magnetically active M sites with $d^1$ electronic configuration are located at the center of octahedral cage. Each cage is formed by X sites with $p^6$ electronic configuration. Indirect hopping paths between M and X sites are marked as dotted curved lines.} 
	\label{fig:kekule}
\end{figure}
	
We first introduce $\alpha-$MX$_3$, a two dimensional edge-shared octahedral system with two types of atoms. The honeycomb geometry is formed by magnetically active cation ions (M) with $d^1$ electronic configuration. Each M site is surrounded by an edge-shared octahedral cage, which is formed by anion ions (X) with $p^6$ electronic configuration, as shown in Fig.~\ref{fig:kekule}. We focus on a case with strong crystal field splitting where one electron resides in threefold degenerate $t_{2g}$ orbitals at each M site. In the $\alpha-$MX$_3$ crystal structure, three distinct nearest-neighbor bonds are formed in $yz$, $zx$, and $xy$ planes, respectively. We label such bonds and the $t_{2g}$ orbitals with an axis $\gamma~(=a,b,c)$ normal to its plane: $a$ indicates $yz$. We first consider the leading hopping process between neighboring $t_{2g}$ orbitals due to direct hybridization which is illustrated as a green bond in Fig.~\ref{fig:kekule}. In $\gamma$ plane, the hopping occurs between $t_{2g}$ orbitals of same $\gamma$ type. The corresponding low energy Hamiltonian for such process reads as the following form in the limit of zero Hund's coupling \cite{jackeli2007dimer, natorij}.
\bea
H_d \! & = & \! J_d \sum_{\langle ij \rangle \parallel \gamma}
\Big( \vec{S}_i \! \cdot \! \vec{S}_j + \frac{1}{4}\Big)n_{i}^{\gamma}n_{j}^{\gamma}
\label{eq:H_d}
\eea
where $\langle ij \rangle \!\! \parallel \!\! \gamma$ indicates $\gamma$ type bond. Here, $\vec{S}_i$ are spin-$\frac{1}{2}$ operators, $n_{i}^{\gamma}$ is the number operator for orbital $\gamma$, and $J_d=4t_{d}^2/U$ where $t_d$ is the amplitude of direct hopping and $U$ is the Coulomb repulsion at a M site.
	
The antiferromagnetic term $H_d$ acts on the $\gamma$ type bonds only when both sites involved are occupied by an electron with $\gamma$ orbital. Accordingly, the antiferromagnetic bonds form non-intersecting linear chains with specific orbital patterns. As the length of the chain grows, a negative energy gain from the antiferromagnetic spin interactions increases. At the same time, each bond gains a positive energy, $\frac{1}{4}$, from the second term in Eq.~\ref{eq:H_d}. Under these two conditions, the minimum energy of the system is achieved when all chains are dimers as described in Ref.~\cite{jackeli2007dimer}. Therefore, a honeycomb lattice covered with hard-core dimers corresponds to an exact ground state of $H_d$. Within a dimer, two electrons occupy the same orbital, which matches the direction of the bond, and form a spin-singlet. The $H_d$ term is inactive on every inter-dimer bond and thus the ground state manifold becomes extensively degenerate: there are infinitely many different ways to cover the lattice with dimers. In the following, we consider an additional hopping process which results in inter-dimer interactions that completely lift the degeneracy and stabilize a valence bond crystal state.
	
In addition to direct hopping between neighboring $t_{2g}$ orbitals, there exist different exchange mechanisms: charge transfer and cyclic exchange processes \cite{liu2022exchange}. We first consider $pd$ charge transfer process, $d^1_i\!-\!p^6\!-\!d^1_j \rightarrow d^2_i\!-\!p^4\!-\!d^2_j$, where two holes are created on an X site, which connects two neighboring M sites as illustrated in Fig.~\ref{fig:kekule} with a blue bond. The relative energy of a virtual state is $2U + 2\Delta - 9 U_p$ where $\Delta$ is $p\!-\!d$ charge transfer gap and $U_p$ is Coulomb repulsion at an X site. Collecting possible charge transfer contributions, we obtain the spin-orbital Hamiltonian
\bea
H_{ct} \!=\! J_{ct} \sum_{\langle ij \rangle \parallel \gamma}
\!\!\Big( \vec{S}_i \! \cdot \! \vec{S}_j - \frac{1}{4}\Big) O^{\gamma}.
\label{eq:H_ct}
\eea
where $O^c\!=\!n_{i}^{a}n_{j}^{b}\!+\!n_{i}^{b}n_{j}^{a}$ for a $c$ type bond (See Section I of SM for details). Here, $J_{ct}=2t_s^2/(2U + 2\Delta - 9 U_p)$ and $t_s$ is the amplitude of hopping, which takes place via intermediate $p$ orbitals. Unlike the charge transfer process, the cyclic exchange process, $d^1\!-\!(p^6,p^6)\!-\!d^1 \rightarrow d^2\!-\!(p^5,p^5)\!-\!d^2$, involves two X sites where a hole is created at each X site in the virtual state as shown in Fig.~\ref{fig:kekule} with a red bond. The cyclic exchange Hamiltonian is expressed as
\bea
H_{ce} \!=\! -J_{ce} \sum_{\langle ij \rangle \parallel \gamma}
\Big( \vec{S}_i \! \cdot \! \vec{S}_j
+ \frac{1}{4} \Big)\tilde{O}^{\gamma}
\label{eq:H_ce}
\eea
where $\tilde{O}^c=\Big( d^{\dagger,b}_{i} d_{i}^{a}d^{\dagger,b}_{j} d_{j}^{a}+d^{\dagger,a}_{i} d_{i}^{b}d^{\dagger,a}_{j}  d_{j}^{b}\Big)$ for a $c$ type bond and $J_{ce}=2t_s^2/(2U+2\Delta-10U_p)$. 
	
\begin{figure}[t!]
	\includegraphics[scale=1.1]{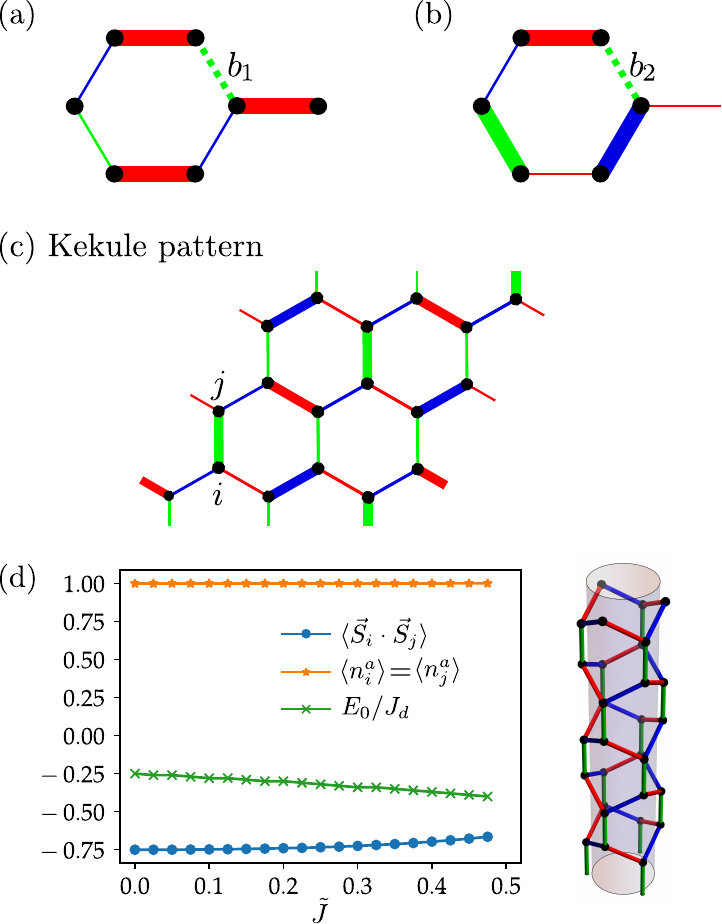}
	\caption{(color online) (a,b) Two different types of inter-dimer bonds, $b_1$ and $b_2$, which are shown as green dashed lines. Each spin-orbital dimer is represented as a thick line and is colored according to the orbital it occupies. $b_1$ couple two dimers with the same orbital. Meanwhile, the bond $b_2$ couple dimers with different orbitals. (c) Kekule dimerized pattern which shows the modification of particular bonds of the honeycomb lattice. (d) Plot of the per site energy $E_0$, spin correlation $\langle \vec{S}_i\cdot\vec{S}_j \rangle$, and electron density, $\langle n_{i}^a \rangle$ and $\langle n_{j}^a \rangle$ for the ground state on YC-6 cylinder. Here, $i$ and $j$ are nearest neighboring sites and connected through $yz$ type bond as shown in (c).} 
	\label{fig:corrs}
\end{figure}
	
$H_{ct}$ generates interactions between dimers and, as it follows, selects a particular superstructure of dimers, a Kekule dimerization pattern. We assume that the charge transfer term is small and treat $H_{ct}$ perturbatively. Figs.~\ref{fig:corrs}(a) and \ref{fig:corrs}(b) contain two different green dotted inter-dimer bonds, $b_1$ and $b_2$, which can be formed in the ground state of $H_d$. $H_{ct}$ is active only on inter-dimer bonds whose two associated dimers have different orbital indices as $b_2$ shown in Fig.~\ref{fig:corrs}(b). At the same time, such bonds gain negative energy, $-\frac{1}{4}$, from the last term in Eq.~(\ref{eq:H_ct}). Accordingly, the system prefers a specific type of dimer orientation, resulting in Kekule dimerized ground state as shown in Fig.~\ref{fig:corrs}(c).
	
To check the stability of the Kekule dimerized state, we have performed infinite density matrix renormalization group (iDMRG) simulation for the Hamiltonian, $H_t \equiv H_d+H_{ct}+H_{ce}$ with $J_{d}\!=\!1$ and $\tilde{J}\!\equiv\!J_{ct}\!=\!2J_{ce}/3$, on YC-2$L_y$ cylinders with circumference $L_y\!=\!2$ and 3, where the periodic boundary condition (PBC) is applied along the $y$ direction (See Section II of SM for details). To find a ground state for each $\tilde{J}$, we compared the per site ground state energy of YC-4 cylinder with the one of YC-6 cylinder. We find that the latter is energetically more stable at least up to $\tilde{J}\!\approx\!1/2$. In Fig.~\ref{fig:corrs}(d), we plot the per site energy $E_0$, nearest-neighbor spin correlation $\langle \vec{S}_i\cdot\vec{S}_j \rangle$, and electron density in the $yz$ orbital, $\langle n_{i}^a \rangle$ and $\langle n_{j}^a \rangle$, for a given dimerized $ij$ bond, which is $yz$ type. It indicates the stabilization of spin-orbit Kekule dimer state and small hybridization of spin-singlet and spin-triplet state within the dimer. Note that the geometry of YC-4 cylinder cannot support the Kekule dimerized state. We also performed exact diagonalization (ED) on a cluster of six sites using PBC. The result of ED also shows stabilization of the Kekule dimer state at least up to $\tilde{J}\!\approx\!1/2$ (See Section III of SM for details). Below, we focus on a regime where direct hopping contributions dominate the system, $\tilde{J}/J_d=1/8$.
	
Having realized the crystallized dimer ground state, we now find the excitation spectrum by introducing auxiliary bosons on each dimer \cite{sachdev1990bond, romhanyi2011effect, nawa2019triplon}. We define the operator $s^{\dagger,a}$ which creates the spin-singlet state $|s\rangle=(|\uparrow_L\downarrow_R\rangle - |\downarrow_L \uparrow_R \rangle)/\sqrt{2}$ and $t^{\dagger,a}_{1}$, $t^{\dagger,a}_{0}$, and $t^{\dagger,a}_{-1}$ which create the spin-triplet states $|t_1\rangle=|\uparrow_L\uparrow_R\rangle$, $|t_0\rangle=(|\uparrow_L\downarrow_R\rangle + |\downarrow_L \uparrow_R \rangle)/\sqrt{2}$, and $|t_{-1} \rangle=|\downarrow_L \downarrow_R \rangle $, respectively, where $a$ labels the orbital index of dimer. Then, spin operators, which act on an electron with orbital $a$, read:
\bea
\nonumber
S^{+}_{i}n_{i}^{a} &=& \frac{t^{\dagger,a}_{1,i} t^{a}_{0,i} + t^{\dagger,a}_{0,i} t^{a}_{-1,i}}{\sqrt{2}}  \pm \frac{s^{\dagger,a}_i t^{a}_{-1,i} - t^{\dagger,a}_{1,i} s^{a}_i}{\sqrt{2}} \;, \\
\nonumber
S^{-}_{i}n_{i}^{a}&=& \frac{t^{\dagger,a}_{-1,i} t^{a}_{0,i} + t^{\dagger,a}_{0,i} t^{a}_{1,i}}{\sqrt{2}}  \mp \frac{s^{\dagger,a}_i t^{a}_{1,i} - t^{\dagger,a}_{-1,i} s^{a}_i}{\sqrt{2}} \;, \\
S^{z}_{i}n_{i}^{a}&=& \frac{t^{\dagger,a}_{1,i} t^{a}_{1,i} - t^{\dagger,a}_{-1,i} t^{a}_{-1,i}}{2}  \pm \frac{s^{\dagger,a}_i t^{a}_{0,i} + t^{\dagger,a}_{0,i} s^{a}_i}{2}
\label{eq:triplon}
\eea
where $i$ indicates the position of the dimer and the upper (lower) sign is for the left (right) spin within the dimer.
	
\begin{figure}[t!]
	\includegraphics[scale=1.15]{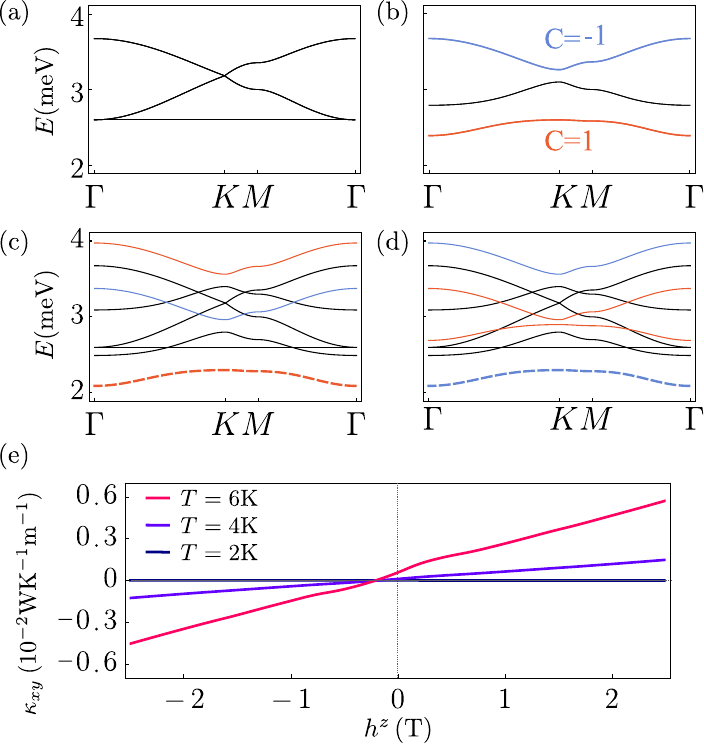}
	\caption{(color online) (a) Complete ($\ket{t_{\pm 1}}$ and $\ket{t_{0}}$) triplon band spectrum of Kekule dimerized state without DMI for $J_d=3$meV. Because of SU(2) spin rotation symmetry, each band is triply degenerate. (b) The band structure of $\ket{t_{1}}$ triplon with $D=0.1$meV. The out–of–plane DMI induce gap and topological property to the band structure. The sign of Chern number, $C$, for each band is opposite for $\ket{t_{-1}}$ triplon. (c,d) Complete triplon band spectrum for $J_d\!=\!3$meV, $D\!=\!0.1$meV, and $h_z\!=\!\pm1$T. The sign of the Chern number for the lowest-dotted band depends on the sign of the external magnetic field $h_z$. (e) Thermal Hall conductivity $\kappa_{xy}$ versus external magnetic field $h_z$ at three distinct temperatures. In the calculation, we include very small in-plane DMI to gap out band touching points so that Berry curvature $F^{(n)}_{xy}(\mathbf{k})$ is well-defined.} 
	\label{fig:band}
\end{figure}

Kekule state is a product of spin-singlet at each dimer, and thus, the density of triplon is zero. So the mean-field theory, which abandons dynamics of singlet operators, is applicable. By replacing these operators with expectation value, $\langle s^{\dagger}_{i} \rangle=\langle s_{i} \rangle=1$, and neglecting high order terms, $H_t$ is rewritten in terms of triplon operators in momentum space as
\bea
\mathcal{H}_\mathbf{k}=\sum_{\mathbf k}
\left(\!\!
\begin{array}{c}
	\mathbf{t}_{1,{\bf k}}^{\dagger}\\
	\mathbf{t}_{1,-\!{\bf k}}^{\phantom{\dagger}}\\
	\mathbf{t}_{0,{\bf k}}^{\dagger}\\
	\mathbf{t}_{0,-\!{\bf k}}^{\phantom{\dagger}}
\end{array}\!
\right)\!\!
\left(\!\!
\begin{array}{cccc}
	M_{1,\mathbf{k}}& N_{1,\mathbf{k}} & \mathbf{0} & \mathbf{0}\\
	N_{1,\mathbf{k}} & M_{1,\mathbf{k}} & \mathbf{0} & \mathbf{0}\\
	\mathbf{0} & \mathbf{0} & M_{0,\mathbf{k}}& N_{0,\mathbf{k}}\\
	\mathbf{0} & \mathbf{0} &  N_{0,\mathbf{k}} & M_{0,\mathbf{k}} 
\end{array}\!\!
\right)
\!\left(\!\!
\begin{array}{c}
	\mathbf{t}_{1,{\bf k}}^{\phantom{\dagger}}\\
	\mathbf{t}_{1,-\!{\bf k}}^{\dagger}\\
	\mathbf{t}_{0,{\bf k}}^{\phantom{\dagger}}\\
	\mathbf{t}_{0,-\!{\bf k}}^{\dagger}
\end{array}\!
\right)\nonumber\\
\label{eq:BdG}
\eea
where $\mathbf{t}_{1,{\bf k}}^{\dagger}=(t_{1,{\bf k}}^{\dagger,a}, t_{{1,{\bf k}}}^{\dagger,b}, t_{{1,{\bf k}}}^{\dagger,c}, t_{{-1,{\bf k}}}^{\dagger,a}, t_{{-1,{\bf k}}}^{\dagger,b}, t_{{-1,{\bf k}}}^{\dagger,c})$ and $\mathbf{t}_{0,{\bf k}}^{\dagger}=(t_{{0,{\bf k}}}^{\dagger,a}, t_{{0,{\bf k}}}^{\dagger,b}, t_{{0,{\bf k}}}^{\dagger,c})$. The Bogoliubov–de Gennes Hamiltonian, $\mathcal{H}_\mathbf{k}$, is separated into two blocks which make manifest the absence of terms mixing spinful $|t_{\pm1}\rangle$ triplon and spinless $|t_0\rangle$ triplon. $\mathcal{H}_\mathbf{k}$ can be conveniently expressed by using 8 Gell-Mann matrices, $\lambda_i$, as the basis for orbital indices $a, b$, and $c$ (See Section IV of SM for the Gell-Mann matrix). The hopping matrices are written as
\bea
\nonumber
M_{1,\mathbf{k}}&\!=\!&J_d [I_2\!\otimes\! I_3] + J_{ct} \big[\cos\frac{\boldsymbol{\delta}_1\cdot\mathbf{k}}{2} I_2\!\otimes \!\lambda_4 \\
\nonumber 
&+&\cos\frac{\boldsymbol{\delta}_2\cdot\mathbf{k}}{2}I_2\!\otimes \!\lambda_1 + \cos\frac{\boldsymbol{\delta}_3\cdot\mathbf{k}}{2} I_2 \!\otimes\! \lambda_6 \big], \\
\nonumber
M_{0,\mathbf{k}}&\!=\!&J_d [I_3] + J_{ct} [\cos\frac{\boldsymbol{\delta}_1\cdot\mathbf{k}}{2}\lambda_4 \\
&+&\cos\frac{\boldsymbol{\delta}_2\cdot\mathbf{k}}{2}\lambda_1 + \cos\frac{\boldsymbol{\delta}_3\cdot\mathbf{k}}{2}\lambda_6]
\eea
where $\boldsymbol{\delta}_1$ and $\boldsymbol{\delta}_2$ are two primitive vectors of the kagome superlattice and $\boldsymbol{\delta}_3\!=\!-\boldsymbol{\delta}_1-\boldsymbol{\delta}_2$. The pairing matrices are given as 
\bea
\nonumber
N_{1,\mathbf{k}}&=&-M_{1,\mathbf{k}}+J_d \left[I_2\!\otimes\! I_3\right],\\
N_{0,\mathbf{k}}&=&M_{0,\mathbf{k}}-J_d \left[I_3\right].
\eea
The dispersion can be obtained by Bogoliubov transformation and the resulting band structure with nine triplon bands is shown in Fig.~\ref{fig:band}(a). As a result of SU(2) spin rotation symmetry, three bands are all three-fold degenerate. At the same time, discrete lattice symmetries give rise to the emergence of quadratic band touching at $\Gamma$ point and linear band crossings at $K$ point resembling the electronic band structure of kagome lattice \cite{bergman2008band}.
	
To investigate the effect of anisotropic terms on the triplon band structure, we include small inter-dimer Dzyaloshinskii-Moriya interactions (DMI), $\sum_{\langle ij \rangle \parallel \gamma} D ( S^x_i S^y_j - S^y_i S^x_j  )O^{\gamma}$, which are allowed by the lattice symmetry. The effect of DMI can be encoded by including $M_{1,\mathbf k}^{D} \!=\! D [\cos\frac{\boldsymbol{\delta}_1\cdot\mathbf{k}}{2} S_z \!\otimes\! \lambda_5+ \cos\frac{\boldsymbol{\delta}_2\cdot\mathbf{k}}{2} S_z\!\otimes \!\lambda_2+ \cos\frac{\boldsymbol{\delta}_3\cdot\mathbf{k}}{2} S_z \!\otimes\! \lambda_7]$ and $N_{1,\mathbf k}^{D} \!=\! -M_{1,\mathbf k}^{D} $ to the spinful hopping matrix $M_{1, \mathbf k}$ and pairing matrix $N_{1, \mathbf k}$ in Eq.~(\ref{eq:BdG}), respectively. The intra-dimer DMI are forbidden since the center of each intra-dimer bond is an inversion center. The out–of–plane DMI still preserve the U(1) spin rotation symmetry and the excitation modes of different triplon, $\ket{t_1}$, $\ket{t_{-1}}$, and $\ket{t_0}$, completely decouple from each other. At the same time, it gap out all the band touching points in $\ket{t_{\pm 1}}$ triplon band structure separately and leave non-trivial topological bands. In Fig.~\ref{fig:band}(b), we plot $\ket{t_{1}}$ triplon band spectrum in the presence of DMI where the bottom (top) band carries the finite Chern number 1 (-1). Although the band spectrum of $\ket{t_{-1}}$ triplon is exactly the same as the one of $\ket{t_{1}}$ tripon, it contains opposite topological property: bottom (top) band carries the Chern number -1 (1). As pointed out in Ref.\cite{thomasen2021fragility}, such an opposite pattern could lead to the triplon analog of $Z_2$ spin-hall insulator state with the corresponding helical edge states at the boundary of the system. However, the $Z_2$ topological phase is destroyed by infinitesimal inter-dimer in-plane DMI, which break the U(1) and consequently pseudo time-reversal symmetry. The pseudo time-reversal operator is defined as U(1)$\times\mathcal{T}$ with physical time reversal operator $\mathcal{T}$. The in-plane DMI is symmetrically allowed in $\alpha-$MX$_3$ where a kagome plane is no longer a mirror plane since the octahedral cage surrounding M ion is tilted \cite{cheong2007multiferroics}. Below, we neglect the in-plane DMI and keep U(1) symmetry intact. 
	
The external magnetic field gives a knob to induce topological transitions through the Zeeman term, $\sum_{i,a} g_z \mu_B h^z S^z_in_{i}^{a}$. Here, $g_z\!\approx\!2$ is the Landé $g$-factor, $\mu_B$ the Bohr magneton, and $h_z$ an applied magnetic field perpendicular to the plane. In Figs.~\ref{fig:band}(c) and \ref{fig:band}(d), we plot the complete triplon band spectrum with finite DMI at two different fields. With $h_z\!>\!0$, the well-separated lowest band is associated to $\ket{t_{-1}}$ triplon and carries the Chern number 1. On the other hand, the lowest band is formed by $\ket{t_{1}}$ triplon with the Chern number -1 when $h_z\!<\!0$.
	
The Chern number in electronic systems is directly connected to physical observable quantities such as quantized Hall conductance. In bosonic systems where it is not possible, the thermal Hall effect gives an alternative \cite{onose2010observation,romhanyi2015hall}. It depends on the Berry curvature in which the bosons are thermally populated, rather than solely depending on the Chern number. In our case, triplons carry a thermal current perpendicular to the temperature gradient when the temperature is above the triplon gap. The thermal Hall conductivity of triplons is written as
\bea
\kappa_{xy} = -\frac{k_{\rm B}^2 T}{\hbar L} \sum_{\mathbf{k},n} c_2[ \rho (\omega_{n,\mathbf{k}}) ] F^{(n)}_{xy}(\mathbf{k})
\label{eq:kxy}
\eea
where $F^{(n)}_{xy}(\mathbf{k})$ is the Berry curvature of $n$th band at $\mathbf{k}$, $c_2(\rho)\!=\!\int_0^\rho \ln^2(1+t^{-1})dt$, and $\rho (\omega_{n,\mathbf{k}})\!=\!(\exp(\omega_{n,\mathbf{k}}/k_{\rm B} T)-1)^{-1}$ is the Bose distribution function. Since $\alpha-$MX$_3$ is quasi two-dimensional material and the thermal conductivity of each layer is in the same direction, we sum the contribution of each layer to calculate the conductivity for a three-dimensional sample with a single layer thickness $L=0.5$nm \cite{jain2013commentary}. Fig.~\ref{fig:band}(e) shows the thermal Hall conductivity $\kappa_{xy}$ as a function of the magnetic field $h_z$. As $h_z\!>\!0$ is turned on, $\kappa_{xy}$ starts to increase. The increase originates from the fact that the well-seperated lowest band, which is formed by $\ket{t_{-1}}$ triplon and most thermally populated, carries the Chern number -1 and dominates the Hall response. On the contrary, $\kappa_{xy}$ decreases when the field $h_z<0$ is included which can be interpreted by the dominance of the lowest $\ket{t_{1}}$ triplon band with the Chern number 1. Such an opposite response can be a direct signature of Kekule dimerized phase in $\alpha-$MX$_3$. In our calculation, we focus on a low temperature regime where triplon bands are weakly populated and neglect interaction effects.
	
In the present work, we demonstrate that the neglected charge transfer contribution crystalizes the system into Kekule spin-orbit dimerized state. To check the stability, we perform iDMRG simulation which indicates a small hybridization of spin-singlet and triplet within each dimer. We then introduce triplon operators and investigate the triplon band structure which is topologically non-trivial in the presence of inter-dimer DMI. For the experimental verification of Kekule dimer phase, we calculate the thermal Hall conductivity, which directly encodes the topological profile of the triplon band spectrum.
	
As a future direction, it would be desirable to apply the degeneracy lifting mechanism to many valence bond crystals with $d^1$ ions which include MgTi$_2$O$_4$ with Ti$^{3+}$ on a pyrochlore lattice \cite{di2004valence}. An essential question regards then whether it can verify the experimentally observed helical dimerization pattern \cite{schmidt2004spin}. Regarding the fact that $d^5$ ions give the exactly same form of spin-orbital Hamiltonian, it would be interesting to study the effect of charge-transfer processes for such systems. Another challenging but interesting avenue is to take into account the triplon interaction effects which might stabilize the topological gap \cite{mook2021interaction} or induce the emergence of triplon bound states \cite{mcclarty2017topological}.
	
\section{Acknowledgments} The author thanks S. B. Lee, M. J. Park, H. Kee, K. Penc, J. Romhányi, and J. Knolle for stimulating conversations, and the latter also for detailed comments on the experimental identification. G.B.S. is funded by the European Research Council (ERC) under the European Unions Horizon 2020 research and innovation program (grant agreement No. 771537). The research is part of the Munich Quantum Valley, which is supported by the Bavarian state government with funds from the Hightech Agenda Bayern Plus. Tensor network calculations were performed using the TeNPy Library \cite{hauschild2018efficient}.

\bibliography{kekule_bib}

\begin{thebibliography}{33}%
\makeatletter
\providecommand \@ifxundefined [1]{%
 \@ifx{#1\undefined}
}%
\providecommand \@ifnum [1]{%
 \ifnum #1\expandafter \@firstoftwo
 \else \expandafter \@secondoftwo
 \fi
}%
\providecommand \@ifx [1]{%
 \ifx #1\expandafter \@firstoftwo
 \else \expandafter \@secondoftwo
 \fi
}%
\providecommand \natexlab [1]{#1}%
\providecommand \enquote  [1]{``#1''}%
\providecommand \bibnamefont  [1]{#1}%
\providecommand \bibfnamefont [1]{#1}%
\providecommand \citenamefont [1]{#1}%
\providecommand \href@noop [0]{\@secondoftwo}%
\providecommand \href [0]{\begingroup \@sanitize@url \@href}%
\providecommand \@href[1]{\@@startlink{#1}\@@href}%
\providecommand \@@href[1]{\endgroup#1\@@endlink}%
\providecommand \@sanitize@url [0]{\catcode `\\12\catcode `\$12\catcode
  `\&12\catcode `\#12\catcode `\^12\catcode `\_12\catcode `\%12\relax}%
\providecommand \@@startlink[1]{}%
\providecommand \@@endlink[0]{}%
\providecommand \url  [0]{\begingroup\@sanitize@url \@url }%
\providecommand \@url [1]{\endgroup\@href {#1}{\urlprefix }}%
\providecommand \urlprefix  [0]{URL }%
\providecommand \Eprint [0]{\href }%
\providecommand \doibase [0]{https://doi.org/}%
\providecommand \selectlanguage [0]{\@gobble}%
\providecommand \bibinfo  [0]{\@secondoftwo}%
\providecommand \bibfield  [0]{\@secondoftwo}%
\providecommand \translation [1]{[#1]}%
\providecommand \BibitemOpen [0]{}%
\providecommand \bibitemStop [0]{}%
\providecommand \bibitemNoStop [0]{.\EOS\space}%
\providecommand \EOS [0]{\spacefactor3000\relax}%
\providecommand \BibitemShut  [1]{\csname bibitem#1\endcsname}%
\let\auto@bib@innerbib\@empty
\bibitem [{\citenamefont {Auerbach}(2012)}]{auerbach2012interacting}%
  \BibitemOpen
  \bibfield  {author} {\bibinfo {author} {\bibfnamefont {A.}~\bibnamefont
  {Auerbach}},\ }\href@noop {} {\emph {\bibinfo {title} {Interacting electrons
  and quantum magnetism}}}\ (\bibinfo  {publisher} {Springer Science \&
  Business Media},\ \bibinfo {year} {2012})\BibitemShut {NoStop}%
\bibitem [{\citenamefont {Henley}(1989)}]{henley1989ordering}%
  \BibitemOpen
  \bibfield  {author} {\bibinfo {author} {\bibfnamefont {C.~L.}\ \bibnamefont
  {Henley}},\ }\href@noop {} {\bibfield  {journal} {\bibinfo  {journal}
  {Physical review letters}\ }\textbf {\bibinfo {volume} {62}},\ \bibinfo
  {pages} {2056} (\bibinfo {year} {1989})}\BibitemShut {NoStop}%
\bibitem [{\citenamefont {Baskaran}\ \emph {et~al.}(2008)\citenamefont
  {Baskaran}, \citenamefont {Sen},\ and\ \citenamefont
  {Shankar}}]{baskaran2008spin}%
  \BibitemOpen
  \bibfield  {author} {\bibinfo {author} {\bibfnamefont {G.}~\bibnamefont
  {Baskaran}}, \bibinfo {author} {\bibfnamefont {D.}~\bibnamefont {Sen}},\ and\
  \bibinfo {author} {\bibfnamefont {R.}~\bibnamefont {Shankar}},\ }\href@noop
  {} {\bibfield  {journal} {\bibinfo  {journal} {Physical Review B}\ }\textbf
  {\bibinfo {volume} {78}},\ \bibinfo {pages} {115116} (\bibinfo {year}
  {2008})}\BibitemShut {NoStop}%
\bibitem [{\citenamefont {Rousochatzakis}\ and\ \citenamefont
  {Perkins}(2017)}]{rousochatzakis2017classical}%
  \BibitemOpen
  \bibfield  {author} {\bibinfo {author} {\bibfnamefont {I.}~\bibnamefont
  {Rousochatzakis}}\ and\ \bibinfo {author} {\bibfnamefont {N.~B.}\
  \bibnamefont {Perkins}},\ }\href@noop {} {\bibfield  {journal} {\bibinfo
  {journal} {Physical review letters}\ }\textbf {\bibinfo {volume} {118}},\
  \bibinfo {pages} {147204} (\bibinfo {year} {2017})}\BibitemShut {NoStop}%
\bibitem [{\citenamefont {Canals}\ and\ \citenamefont
  {Lacroix}(1998)}]{canals1998pyrochlore}%
  \BibitemOpen
  \bibfield  {author} {\bibinfo {author} {\bibfnamefont {B.}~\bibnamefont
  {Canals}}\ and\ \bibinfo {author} {\bibfnamefont {C.}~\bibnamefont
  {Lacroix}},\ }\href@noop {} {\bibfield  {journal} {\bibinfo  {journal}
  {Physical Review Letters}\ }\textbf {\bibinfo {volume} {80}},\ \bibinfo
  {pages} {2933} (\bibinfo {year} {1998})}\BibitemShut {NoStop}%
\bibitem [{\citenamefont {Baskaran}\ \emph {et~al.}(2007)\citenamefont
  {Baskaran}, \citenamefont {Mandal},\ and\ \citenamefont
  {Shankar}}]{baskaran2007exact}%
  \BibitemOpen
  \bibfield  {author} {\bibinfo {author} {\bibfnamefont {G.}~\bibnamefont
  {Baskaran}}, \bibinfo {author} {\bibfnamefont {S.}~\bibnamefont {Mandal}},\
  and\ \bibinfo {author} {\bibfnamefont {R.}~\bibnamefont {Shankar}},\
  }\href@noop {} {\bibfield  {journal} {\bibinfo  {journal} {Physical review
  letters}\ }\textbf {\bibinfo {volume} {98}},\ \bibinfo {pages} {247201}
  (\bibinfo {year} {2007})}\BibitemShut {NoStop}%
\bibitem [{\citenamefont {Bergman}\ \emph {et~al.}(2007)\citenamefont
  {Bergman}, \citenamefont {Alicea}, \citenamefont {Gull}, \citenamefont
  {Trebst},\ and\ \citenamefont {Balents}}]{bergman2007order}%
  \BibitemOpen
  \bibfield  {author} {\bibinfo {author} {\bibfnamefont {D.}~\bibnamefont
  {Bergman}}, \bibinfo {author} {\bibfnamefont {J.}~\bibnamefont {Alicea}},
  \bibinfo {author} {\bibfnamefont {E.}~\bibnamefont {Gull}}, \bibinfo {author}
  {\bibfnamefont {S.}~\bibnamefont {Trebst}},\ and\ \bibinfo {author}
  {\bibfnamefont {L.}~\bibnamefont {Balents}},\ }\href@noop {} {\bibfield
  {journal} {\bibinfo  {journal} {Nature Physics}\ }\textbf {\bibinfo {volume}
  {3}},\ \bibinfo {pages} {487} (\bibinfo {year} {2007})}\BibitemShut {NoStop}%
\bibitem [{\citenamefont {Lee}\ \emph {et~al.}(2014)\citenamefont {Lee},
  \citenamefont {Lee}, \citenamefont {Paramekanti},\ and\ \citenamefont
  {Kim}}]{lee2014order}%
  \BibitemOpen
  \bibfield  {author} {\bibinfo {author} {\bibfnamefont {S.}~\bibnamefont
  {Lee}}, \bibinfo {author} {\bibfnamefont {E.~K.-H.}\ \bibnamefont {Lee}},
  \bibinfo {author} {\bibfnamefont {A.}~\bibnamefont {Paramekanti}},\ and\
  \bibinfo {author} {\bibfnamefont {Y.~B.}\ \bibnamefont {Kim}},\ }\href@noop
  {} {\bibfield  {journal} {\bibinfo  {journal} {Physical Review B}\ }\textbf
  {\bibinfo {volume} {89}},\ \bibinfo {pages} {014424} (\bibinfo {year}
  {2014})}\BibitemShut {NoStop}%
\bibitem [{\citenamefont {Yamashita}\ and\ \citenamefont
  {Ueda}(2000)}]{yamashita2000spin}%
  \BibitemOpen
  \bibfield  {author} {\bibinfo {author} {\bibfnamefont {Y.}~\bibnamefont
  {Yamashita}}\ and\ \bibinfo {author} {\bibfnamefont {K.}~\bibnamefont
  {Ueda}},\ }\href@noop {} {\bibfield  {journal} {\bibinfo  {journal} {Physical
  review letters}\ }\textbf {\bibinfo {volume} {85}},\ \bibinfo {pages} {4960}
  (\bibinfo {year} {2000})}\BibitemShut {NoStop}%
\bibitem [{\citenamefont {Tchernyshyov}\ \emph {et~al.}(2002)\citenamefont
  {Tchernyshyov}, \citenamefont {Moessner},\ and\ \citenamefont
  {Sondhi}}]{tchernyshyov2002order}%
  \BibitemOpen
  \bibfield  {author} {\bibinfo {author} {\bibfnamefont {O.}~\bibnamefont
  {Tchernyshyov}}, \bibinfo {author} {\bibfnamefont {R.}~\bibnamefont
  {Moessner}},\ and\ \bibinfo {author} {\bibfnamefont {S.}~\bibnamefont
  {Sondhi}},\ }\href@noop {} {\bibfield  {journal} {\bibinfo  {journal}
  {Physical review letters}\ }\textbf {\bibinfo {volume} {88}},\ \bibinfo
  {pages} {067203} (\bibinfo {year} {2002})}\BibitemShut {NoStop}%
\bibitem [{\citenamefont {Jackeli}\ and\ \citenamefont
  {Ivanov}(2007)}]{jackeli2007dimer}%
  \BibitemOpen
  \bibfield  {author} {\bibinfo {author} {\bibfnamefont {G.}~\bibnamefont
  {Jackeli}}\ and\ \bibinfo {author} {\bibfnamefont {D.}~\bibnamefont
  {Ivanov}},\ }\href@noop {} {\bibfield  {journal} {\bibinfo  {journal}
  {Physical Review B}\ }\textbf {\bibinfo {volume} {76}},\ \bibinfo {pages}
  {132407} (\bibinfo {year} {2007})}\BibitemShut {NoStop}%
\bibitem [{\citenamefont {Bastien}\ \emph {et~al.}(2018)\citenamefont
  {Bastien}, \citenamefont {Garbarino}, \citenamefont {Yadav}, \citenamefont
  {Martinez-Casado}, \citenamefont {Rodr{\'\i}guez}, \citenamefont {Stahl},
  \citenamefont {Kusch}, \citenamefont {Limandri}, \citenamefont {Ray},
  \citenamefont {Lampen-Kelley} \emph {et~al.}}]{bastien2018pressure}%
  \BibitemOpen
  \bibfield  {author} {\bibinfo {author} {\bibfnamefont {G.}~\bibnamefont
  {Bastien}}, \bibinfo {author} {\bibfnamefont {G.}~\bibnamefont {Garbarino}},
  \bibinfo {author} {\bibfnamefont {R.}~\bibnamefont {Yadav}}, \bibinfo
  {author} {\bibfnamefont {F.~J.}\ \bibnamefont {Martinez-Casado}}, \bibinfo
  {author} {\bibfnamefont {R.~B.}\ \bibnamefont {Rodr{\'\i}guez}}, \bibinfo
  {author} {\bibfnamefont {Q.}~\bibnamefont {Stahl}}, \bibinfo {author}
  {\bibfnamefont {M.}~\bibnamefont {Kusch}}, \bibinfo {author} {\bibfnamefont
  {S.}~\bibnamefont {Limandri}}, \bibinfo {author} {\bibfnamefont
  {R.}~\bibnamefont {Ray}}, \bibinfo {author} {\bibfnamefont {P.}~\bibnamefont
  {Lampen-Kelley}}, \emph {et~al.},\ }\href@noop {} {\bibfield  {journal}
  {\bibinfo  {journal} {Physical Review B}\ }\textbf {\bibinfo {volume} {97}},\
  \bibinfo {pages} {241108} (\bibinfo {year} {2018})}\BibitemShut {NoStop}%
\bibitem [{\citenamefont {de~Vries}\ \emph {et~al.}(2010)\citenamefont
  {de~Vries}, \citenamefont {Mclaughlin},\ and\ \citenamefont
  {Bos}}]{de2010valence}%
  \BibitemOpen
  \bibfield  {author} {\bibinfo {author} {\bibfnamefont {M.~A.}\ \bibnamefont
  {de~Vries}}, \bibinfo {author} {\bibfnamefont {A.}~\bibnamefont
  {Mclaughlin}},\ and\ \bibinfo {author} {\bibfnamefont {J.-W.}\ \bibnamefont
  {Bos}},\ }\href@noop {} {\bibfield  {journal} {\bibinfo  {journal} {Physical
  review letters}\ }\textbf {\bibinfo {volume} {104}},\ \bibinfo {pages}
  {177202} (\bibinfo {year} {2010})}\BibitemShut {NoStop}%
\bibitem [{\citenamefont {Romh{\'a}nyi}\ \emph {et~al.}(2017)\citenamefont
  {Romh{\'a}nyi}, \citenamefont {Balents},\ and\ \citenamefont
  {Jackeli}}]{romhanyi2017spin}%
  \BibitemOpen
  \bibfield  {author} {\bibinfo {author} {\bibfnamefont {J.}~\bibnamefont
  {Romh{\'a}nyi}}, \bibinfo {author} {\bibfnamefont {L.}~\bibnamefont
  {Balents}},\ and\ \bibinfo {author} {\bibfnamefont {G.}~\bibnamefont
  {Jackeli}},\ }\href@noop {} {\bibfield  {journal} {\bibinfo  {journal}
  {Physical review letters}\ }\textbf {\bibinfo {volume} {118}},\ \bibinfo
  {pages} {217202} (\bibinfo {year} {2017})}\BibitemShut {NoStop}%
\bibitem [{\citenamefont {Miura}\ \emph {et~al.}(2007)\citenamefont {Miura},
  \citenamefont {Yasui}, \citenamefont {Sato}, \citenamefont {Igawa},\ and\
  \citenamefont {Kakurai}}]{miura2007new}%
  \BibitemOpen
  \bibfield  {author} {\bibinfo {author} {\bibfnamefont {Y.}~\bibnamefont
  {Miura}}, \bibinfo {author} {\bibfnamefont {Y.}~\bibnamefont {Yasui}},
  \bibinfo {author} {\bibfnamefont {M.}~\bibnamefont {Sato}}, \bibinfo {author}
  {\bibfnamefont {N.}~\bibnamefont {Igawa}},\ and\ \bibinfo {author}
  {\bibfnamefont {K.}~\bibnamefont {Kakurai}},\ }\href@noop {} {\bibfield
  {journal} {\bibinfo  {journal} {Journal of the Physical Society of Japan}\
  }\textbf {\bibinfo {volume} {76}},\ \bibinfo {pages} {033705} (\bibinfo
  {year} {2007})}\BibitemShut {NoStop}%
\bibitem [{\citenamefont {Jackeli}\ and\ \citenamefont
  {Khomskii}(2008)}]{jackeli2008classical}%
  \BibitemOpen
  \bibfield  {author} {\bibinfo {author} {\bibfnamefont {G.}~\bibnamefont
  {Jackeli}}\ and\ \bibinfo {author} {\bibfnamefont {D.}~\bibnamefont
  {Khomskii}},\ }\href@noop {} {\bibfield  {journal} {\bibinfo  {journal}
  {Physical review letters}\ }\textbf {\bibinfo {volume} {100}},\ \bibinfo
  {pages} {147203} (\bibinfo {year} {2008})}\BibitemShut {NoStop}%
\bibitem [{\citenamefont {B.}\ and\ \citenamefont
  {Flengas}(1964)}]{b1964synthesis}%
  \BibitemOpen
  \bibfield  {author} {\bibinfo {author} {\bibnamefont {B.}}\ and\ \bibinfo
  {author} {\bibfnamefont {S.}~\bibnamefont {Flengas}},\ }\href@noop {}
  {\bibfield  {journal} {\bibinfo  {journal} {Canadian Journal of Chemistry}\
  }\textbf {\bibinfo {volume} {42}},\ \bibinfo {pages} {1495} (\bibinfo {year}
  {1964})}\BibitemShut {NoStop}%
\bibitem [{\citenamefont {Natori}()}]{natorij}%
  \BibitemOpen
  \bibfield  {author} {\bibinfo {author} {\bibfnamefont {W.~M.~H.}\
  \bibnamefont {Natori}},\ }\emph {\bibinfo {title} {j= 3/2 Quantum
  spin-orbital liquids}},\ \href@noop {} {Ph.D. thesis},\ \bibinfo  {school}
  {Universidade de S{\~a}o Paulo}\BibitemShut {NoStop}%
\bibitem [{\citenamefont {Liu}\ \emph {et~al.}(2022)\citenamefont {Liu},
  \citenamefont {Chaloupka},\ and\ \citenamefont
  {Khaliullin}}]{liu2022exchange}%
  \BibitemOpen
  \bibfield  {author} {\bibinfo {author} {\bibfnamefont {H.}~\bibnamefont
  {Liu}}, \bibinfo {author} {\bibfnamefont {J.}~\bibnamefont {Chaloupka}},\
  and\ \bibinfo {author} {\bibfnamefont {G.}~\bibnamefont {Khaliullin}},\
  }\href@noop {} {\bibfield  {journal} {\bibinfo  {journal} {Physical Review
  B}\ }\textbf {\bibinfo {volume} {105}},\ \bibinfo {pages} {214411} (\bibinfo
  {year} {2022})}\BibitemShut {NoStop}%
\bibitem [{\citenamefont {Sachdev}\ and\ \citenamefont
  {Bhatt}(1990)}]{sachdev1990bond}%
  \BibitemOpen
  \bibfield  {author} {\bibinfo {author} {\bibfnamefont {S.}~\bibnamefont
  {Sachdev}}\ and\ \bibinfo {author} {\bibfnamefont {R.}~\bibnamefont
  {Bhatt}},\ }\href@noop {} {\bibfield  {journal} {\bibinfo  {journal}
  {Physical Review B}\ }\textbf {\bibinfo {volume} {41}},\ \bibinfo {pages}
  {9323} (\bibinfo {year} {1990})}\BibitemShut {NoStop}%
\bibitem [{\citenamefont {Romh{\'a}nyi}\ \emph {et~al.}(2011)\citenamefont
  {Romh{\'a}nyi}, \citenamefont {Totsuka},\ and\ \citenamefont
  {Penc}}]{romhanyi2011effect}%
  \BibitemOpen
  \bibfield  {author} {\bibinfo {author} {\bibfnamefont {J.}~\bibnamefont
  {Romh{\'a}nyi}}, \bibinfo {author} {\bibfnamefont {K.}~\bibnamefont
  {Totsuka}},\ and\ \bibinfo {author} {\bibfnamefont {K.}~\bibnamefont
  {Penc}},\ }\href@noop {} {\bibfield  {journal} {\bibinfo  {journal} {Physical
  Review B}\ }\textbf {\bibinfo {volume} {83}},\ \bibinfo {pages} {024413}
  (\bibinfo {year} {2011})}\BibitemShut {NoStop}%
\bibitem [{\citenamefont {Nawa}\ \emph {et~al.}(2019)\citenamefont {Nawa},
  \citenamefont {Tanaka}, \citenamefont {Kurita}, \citenamefont {Sato},
  \citenamefont {Sugiyama}, \citenamefont {Uekusa}, \citenamefont
  {Ohira-Kawamura}, \citenamefont {Nakajima},\ and\ \citenamefont
  {Tanaka}}]{nawa2019triplon}%
  \BibitemOpen
  \bibfield  {author} {\bibinfo {author} {\bibfnamefont {K.}~\bibnamefont
  {Nawa}}, \bibinfo {author} {\bibfnamefont {K.}~\bibnamefont {Tanaka}},
  \bibinfo {author} {\bibfnamefont {N.}~\bibnamefont {Kurita}}, \bibinfo
  {author} {\bibfnamefont {T.~J.}\ \bibnamefont {Sato}}, \bibinfo {author}
  {\bibfnamefont {H.}~\bibnamefont {Sugiyama}}, \bibinfo {author}
  {\bibfnamefont {H.}~\bibnamefont {Uekusa}}, \bibinfo {author} {\bibfnamefont
  {S.}~\bibnamefont {Ohira-Kawamura}}, \bibinfo {author} {\bibfnamefont
  {K.}~\bibnamefont {Nakajima}},\ and\ \bibinfo {author} {\bibfnamefont
  {H.}~\bibnamefont {Tanaka}},\ }\href@noop {} {\bibfield  {journal} {\bibinfo
  {journal} {Nature communications}\ }\textbf {\bibinfo {volume} {10}},\
  \bibinfo {pages} {1} (\bibinfo {year} {2019})}\BibitemShut {NoStop}%
\bibitem [{\citenamefont {Bergman}\ \emph {et~al.}(2008)\citenamefont
  {Bergman}, \citenamefont {Wu},\ and\ \citenamefont
  {Balents}}]{bergman2008band}%
  \BibitemOpen
  \bibfield  {author} {\bibinfo {author} {\bibfnamefont {D.~L.}\ \bibnamefont
  {Bergman}}, \bibinfo {author} {\bibfnamefont {C.}~\bibnamefont {Wu}},\ and\
  \bibinfo {author} {\bibfnamefont {L.}~\bibnamefont {Balents}},\ }\href@noop
  {} {\bibfield  {journal} {\bibinfo  {journal} {Physical Review B}\ }\textbf
  {\bibinfo {volume} {78}},\ \bibinfo {pages} {125104} (\bibinfo {year}
  {2008})}\BibitemShut {NoStop}%
\bibitem [{\citenamefont {Thomasen}\ \emph {et~al.}(2021)\citenamefont
  {Thomasen}, \citenamefont {Penc}, \citenamefont {Shannon},\ and\
  \citenamefont {Romh{\'a}nyi}}]{thomasen2021fragility}%
  \BibitemOpen
  \bibfield  {author} {\bibinfo {author} {\bibfnamefont {A.}~\bibnamefont
  {Thomasen}}, \bibinfo {author} {\bibfnamefont {K.}~\bibnamefont {Penc}},
  \bibinfo {author} {\bibfnamefont {N.}~\bibnamefont {Shannon}},\ and\ \bibinfo
  {author} {\bibfnamefont {J.}~\bibnamefont {Romh{\'a}nyi}},\ }\href@noop {}
  {\bibfield  {journal} {\bibinfo  {journal} {Physical Review B}\ }\textbf
  {\bibinfo {volume} {104}},\ \bibinfo {pages} {104412} (\bibinfo {year}
  {2021})}\BibitemShut {NoStop}%
\bibitem [{\citenamefont {Cheong}\ and\ \citenamefont
  {Mostovoy}(2007)}]{cheong2007multiferroics}%
  \BibitemOpen
  \bibfield  {author} {\bibinfo {author} {\bibfnamefont {S.-W.}\ \bibnamefont
  {Cheong}}\ and\ \bibinfo {author} {\bibfnamefont {M.}~\bibnamefont
  {Mostovoy}},\ }\href@noop {} {\bibfield  {journal} {\bibinfo  {journal}
  {Nature materials}\ }\textbf {\bibinfo {volume} {6}},\ \bibinfo {pages} {13}
  (\bibinfo {year} {2007})}\BibitemShut {NoStop}%
\bibitem [{\citenamefont {Onose}\ \emph {et~al.}(2010)\citenamefont {Onose},
  \citenamefont {Ideue}, \citenamefont {Katsura}, \citenamefont {Shiomi},
  \citenamefont {Nagaosa},\ and\ \citenamefont
  {Tokura}}]{onose2010observation}%
  \BibitemOpen
  \bibfield  {author} {\bibinfo {author} {\bibfnamefont {Y.}~\bibnamefont
  {Onose}}, \bibinfo {author} {\bibfnamefont {T.}~\bibnamefont {Ideue}},
  \bibinfo {author} {\bibfnamefont {H.}~\bibnamefont {Katsura}}, \bibinfo
  {author} {\bibfnamefont {Y.}~\bibnamefont {Shiomi}}, \bibinfo {author}
  {\bibfnamefont {N.}~\bibnamefont {Nagaosa}},\ and\ \bibinfo {author}
  {\bibfnamefont {Y.}~\bibnamefont {Tokura}},\ }\href@noop {} {\bibfield
  {journal} {\bibinfo  {journal} {Science}\ }\textbf {\bibinfo {volume}
  {329}},\ \bibinfo {pages} {297} (\bibinfo {year} {2010})}\BibitemShut
  {NoStop}%
\bibitem [{\citenamefont {Romh{\'a}nyi}\ \emph {et~al.}(2015)\citenamefont
  {Romh{\'a}nyi}, \citenamefont {Penc},\ and\ \citenamefont
  {Ganesh}}]{romhanyi2015hall}%
  \BibitemOpen
  \bibfield  {author} {\bibinfo {author} {\bibfnamefont {J.}~\bibnamefont
  {Romh{\'a}nyi}}, \bibinfo {author} {\bibfnamefont {K.}~\bibnamefont {Penc}},\
  and\ \bibinfo {author} {\bibfnamefont {R.}~\bibnamefont {Ganesh}},\
  }\href@noop {} {\bibfield  {journal} {\bibinfo  {journal} {Nature
  communications}\ }\textbf {\bibinfo {volume} {6}},\ \bibinfo {pages} {1}
  (\bibinfo {year} {2015})}\BibitemShut {NoStop}%
\bibitem [{\citenamefont {Jain}\ \emph {et~al.}(2013)\citenamefont {Jain},
  \citenamefont {Ong}, \citenamefont {Hautier}, \citenamefont {Chen},
  \citenamefont {Richards}, \citenamefont {Dacek}, \citenamefont {Cholia},
  \citenamefont {Gunter}, \citenamefont {Skinner}, \citenamefont {Ceder} \emph
  {et~al.}}]{jain2013commentary}%
  \BibitemOpen
  \bibfield  {author} {\bibinfo {author} {\bibfnamefont {A.}~\bibnamefont
  {Jain}}, \bibinfo {author} {\bibfnamefont {S.~P.}\ \bibnamefont {Ong}},
  \bibinfo {author} {\bibfnamefont {G.}~\bibnamefont {Hautier}}, \bibinfo
  {author} {\bibfnamefont {W.}~\bibnamefont {Chen}}, \bibinfo {author}
  {\bibfnamefont {W.~D.}\ \bibnamefont {Richards}}, \bibinfo {author}
  {\bibfnamefont {S.}~\bibnamefont {Dacek}}, \bibinfo {author} {\bibfnamefont
  {S.}~\bibnamefont {Cholia}}, \bibinfo {author} {\bibfnamefont
  {D.}~\bibnamefont {Gunter}}, \bibinfo {author} {\bibfnamefont
  {D.}~\bibnamefont {Skinner}}, \bibinfo {author} {\bibfnamefont
  {G.}~\bibnamefont {Ceder}}, \emph {et~al.},\ }\href@noop {} {\bibfield
  {journal} {\bibinfo  {journal} {APL materials}\ }\textbf {\bibinfo {volume}
  {1}},\ \bibinfo {pages} {011002} (\bibinfo {year} {2013})}\BibitemShut
  {NoStop}%
\bibitem [{\citenamefont {Di~Matteo}\ \emph {et~al.}(2004)\citenamefont
  {Di~Matteo}, \citenamefont {Jackeli}, \citenamefont {Lacroix},\ and\
  \citenamefont {Perkins}}]{di2004valence}%
  \BibitemOpen
  \bibfield  {author} {\bibinfo {author} {\bibfnamefont {S.}~\bibnamefont
  {Di~Matteo}}, \bibinfo {author} {\bibfnamefont {G.}~\bibnamefont {Jackeli}},
  \bibinfo {author} {\bibfnamefont {C.}~\bibnamefont {Lacroix}},\ and\ \bibinfo
  {author} {\bibfnamefont {N.}~\bibnamefont {Perkins}},\ }\href@noop {}
  {\bibfield  {journal} {\bibinfo  {journal} {Physical review letters}\
  }\textbf {\bibinfo {volume} {93}},\ \bibinfo {pages} {077208} (\bibinfo
  {year} {2004})}\BibitemShut {NoStop}%
\bibitem [{\citenamefont {Schmidt}\ \emph {et~al.}(2004)\citenamefont
  {Schmidt}, \citenamefont {W~Ratcliff}, \citenamefont {Radaelli},
  \citenamefont {Refson}, \citenamefont {Harrison},\ and\ \citenamefont
  {Cheong}}]{schmidt2004spin}%
  \BibitemOpen
  \bibfield  {author} {\bibinfo {author} {\bibfnamefont {M.}~\bibnamefont
  {Schmidt}}, \bibinfo {author} {\bibfnamefont {I.}~\bibnamefont {W~Ratcliff}},
  \bibinfo {author} {\bibfnamefont {P.}~\bibnamefont {Radaelli}}, \bibinfo
  {author} {\bibfnamefont {K.}~\bibnamefont {Refson}}, \bibinfo {author}
  {\bibfnamefont {N.}~\bibnamefont {Harrison}},\ and\ \bibinfo {author}
  {\bibfnamefont {S.-W.}\ \bibnamefont {Cheong}},\ }\href@noop {} {\bibfield
  {journal} {\bibinfo  {journal} {Physical review letters}\ }\textbf {\bibinfo
  {volume} {92}},\ \bibinfo {pages} {056402} (\bibinfo {year}
  {2004})}\BibitemShut {NoStop}%
\bibitem [{\citenamefont {Mook}\ \emph {et~al.}(2021)\citenamefont {Mook},
  \citenamefont {Plekhanov}, \citenamefont {Klinovaja},\ and\ \citenamefont
  {Loss}}]{mook2021interaction}%
  \BibitemOpen
  \bibfield  {author} {\bibinfo {author} {\bibfnamefont {A.}~\bibnamefont
  {Mook}}, \bibinfo {author} {\bibfnamefont {K.}~\bibnamefont {Plekhanov}},
  \bibinfo {author} {\bibfnamefont {J.}~\bibnamefont {Klinovaja}},\ and\
  \bibinfo {author} {\bibfnamefont {D.}~\bibnamefont {Loss}},\ }\href@noop {}
  {\bibfield  {journal} {\bibinfo  {journal} {Physical Review X}\ }\textbf
  {\bibinfo {volume} {11}},\ \bibinfo {pages} {021061} (\bibinfo {year}
  {2021})}\BibitemShut {NoStop}%
\bibitem [{\citenamefont {McClarty}\ \emph {et~al.}(2017)\citenamefont
  {McClarty}, \citenamefont {Kr{\"u}ger}, \citenamefont {Guidi}, \citenamefont
  {Parker}, \citenamefont {Refson}, \citenamefont {Parker}, \citenamefont
  {Prabhakaran},\ and\ \citenamefont {Coldea}}]{mcclarty2017topological}%
  \BibitemOpen
  \bibfield  {author} {\bibinfo {author} {\bibfnamefont {P.~A.}\ \bibnamefont
  {McClarty}}, \bibinfo {author} {\bibfnamefont {F.}~\bibnamefont
  {Kr{\"u}ger}}, \bibinfo {author} {\bibfnamefont {T.}~\bibnamefont {Guidi}},
  \bibinfo {author} {\bibfnamefont {S.}~\bibnamefont {Parker}}, \bibinfo
  {author} {\bibfnamefont {K.}~\bibnamefont {Refson}}, \bibinfo {author}
  {\bibfnamefont {A.}~\bibnamefont {Parker}}, \bibinfo {author} {\bibfnamefont
  {D.}~\bibnamefont {Prabhakaran}},\ and\ \bibinfo {author} {\bibfnamefont
  {R.}~\bibnamefont {Coldea}},\ }\href@noop {} {\bibfield  {journal} {\bibinfo
  {journal} {Nature Physics}\ }\textbf {\bibinfo {volume} {13}},\ \bibinfo
  {pages} {736} (\bibinfo {year} {2017})}\BibitemShut {NoStop}%
\bibitem [{\citenamefont {Hauschild}\ and\ \citenamefont
  {Pollmann}(2018)}]{hauschild2018efficient}%
  \BibitemOpen
  \bibfield  {author} {\bibinfo {author} {\bibfnamefont {J.}~\bibnamefont
  {Hauschild}}\ and\ \bibinfo {author} {\bibfnamefont {F.}~\bibnamefont
  {Pollmann}},\ }\href@noop {} {\bibfield  {journal} {\bibinfo  {journal}
  {SciPost Physics Lecture Notes}\ ,\ \bibinfo {pages} {005}} (\bibinfo {year}
  {2018})}\BibitemShut {NoStop}%
\end{thebibliography}%
\renewcommand{\thefigure}{S\arabic{figure}}
\setcounter{figure}{0}
\renewcommand{\theequation}{S\arabic{equation}}
\setcounter{equation}{0}

\begin{widetext}

\section{Supplementary Material}
		
\subsection{I. Derivation of spin-orbital model for charge-transfer processes}
		
Here, we consider the charge-transfer processes of the type $d^1_i\!-\!p^6\!-\!d^1_j \rightarrow d^2_i\!-\!p^4\!-\!d^2_j$ where two holes are created on an anion ion connecting nearest-neighbor cation ions $i$ and $j$. The virtual-state energy is $E_v \equiv 2U + 2\Delta - 9 U_p$. Here we give details about the derivation for bonds on the $xy$ planes for concreteness. There are four possible electron hopping processes for this case :
\bea
\label{eq:ct2_1}
&&d_{yz}^1\!-\!p_{z}^2\!-\!d_{zx}^1 \rightarrow d_{yz}^2\!-\!p_{z}^0\!-\!d_{zx}^2, \text{ as given in Fig.~\ref{fig:ct3}} \\
\label{eq:ct2_2}
&&d_{yz}^1\!-\!p_{z}^2p_{y}^2\!-\!d_{xy}^1 \rightarrow d_{yz}^2\!-\!p_{z}^1p_{y}^1\!-\!d_{xy}^2, \\
\label{eq:ct2_3}
&&d_{xy}^1\!-\!p_{x}^2p_{z}^2\!-\!d_{zx}^1 \rightarrow d_{xy}^2\!-\!p_{z}^1p_{x}^1\!-\!d_{zx}^2, \\
\label{eq:ct2_4}
&&d_{xy}^1\!-\!p_{x}^2p_{y}^2\!-\!d_{xy}^1 \rightarrow d_{xy}^2\!-\!p_{x}^1p_{y}^1\!-\!d_{xy}^2.
\eea
		
\begin{figure}[h!]
	\includegraphics[width=0.3\columnwidth]{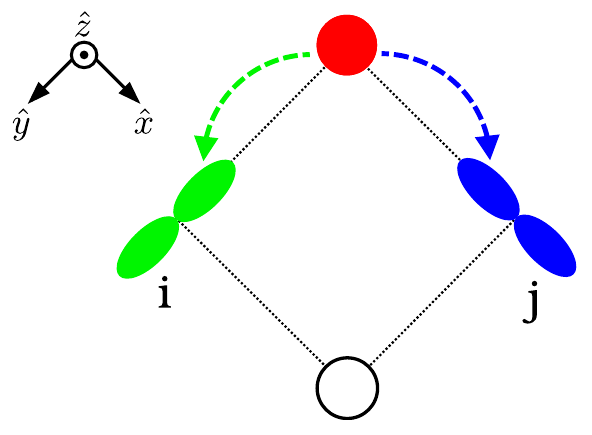}
	\caption{A 90$^{\circ}$ M$_{i}$-X-M$_{j}$ geometry where magnetic ions M$_i$ and M$_j$ interact via an ion X. Here, two holes with opposite spins arrive at the same $p_z$ orbital in an ion X.}
	\label{fig:ct3}
\end{figure}
		
First, we focus on a case with Eq.~(\ref{eq:ct2_1}) and give details about the derivation of effective low-energy spin-orbital Hamiltonian within the subspace where an initial state occupies an electron at site $i (j)$ with $yz (zx)$ orbital. The basis for the subspace is written as
\bea
S_{yz,zx}=\{\ket{yz_{\uparrow},zx_{\uparrow}},\ket{yz_{\uparrow},zx_{\downarrow}},\ket{yz_{\downarrow},zx_{\uparrow}},\ket{yz_{\downarrow},zx_{\downarrow}}\}.
\eea
There are two possible hopping processes: the spin-conserving process and the spin-flipping process. The contribution from the first one can be represented in $S_{yz,zx}$ subspace as
\bea
\big[\bs{H}_{ct}^z\big]_{yz,zx} =-\frac{t_s^2}{E_v}
\left(\!
\begin{array}{cccc}
	0 & 0 & 0 & 0 \\
	0 & 1 & 0 & 0 \\
	0 & 0 & 1 & 0 \\
	0 & 0 & 0 & 0\\
\end{array}
\! \right).
\label{eq:spin-c}
\eea
On the other hand, the spin-flipping process can be represented in $S_{yz,zx}$ subspace as
\bea
\big[\bs{H}_{ct}^z\big]_{yz,zx} =-\frac{t_s^2}{E_v}
\left(\!
\begin{array}{cccc}
	0 & 0 & 0 & 0 \\
	0 & 0 & -1 & 0 \\
	0 & -1 & 0 & 0 \\
	0 & 0 & 0 & 0\\
\end{array}
\! \right).
\label{eq:spin-f}
\eea
Here, initial states $\ket{yz_{\uparrow},zx_{\uparrow}}$ and $\ket{yz_{\downarrow},zx_{\downarrow}}$ cannot give any contribution since two holes with same spins cannot occupy the same orbital at an anion ion. Next, we focus on a case where an initial state occupies an electron at site $i (j)$ with $xy (xy)$ orbital. The basis for such subspace is written as
\bea
S_{xy,xy}=\{\ket{xy_{\uparrow},xy_{\uparrow}},\ket{xy_{\uparrow},xy_{\downarrow}},\ket{xy_{\downarrow},xy_{\uparrow}},\ket{xy_{\downarrow},xy_{\downarrow}}\}.
\eea
Here, the spin-flipping process is not possible and the contribution can be represented in $S_{xy,xy}$ subspace as
\bea
\big[\bs{H}_{ct}^z\big]_{xy,xy}=-\frac{2t_s^2}{E_v}\mathbb{I}_{4 \times 4}.
\eea
The contributions associated to Eq.~(\ref{eq:ct2_1}) in other subspaces are written as
\bea
\nonumber
\big[\bs{H}_{ct}^z\big]_{xy,yz}&=&-\frac{2t_s^2}{E_v}\mathbb{I}_{4 \times 4},~
\big[\bs{H}_{ct}^z\big]_{xy,zx}=-\frac{2t_s^2}{E_v}\mathbb{I}_{4 \times 4},~
\big[\bs{H}_{ct}^z\big]_{yz,xy}=-\frac{t_s^2}{E_v}\mathbb{I}_{4 \times 4},~
\big[\bs{H}_{ct}^z\big]_{yz,yz}=-\frac{t_s^2}{E_v}\mathbb{I}_{4 \times 4},\\
\big[\bs{H}_{ct}^z\big]_{zx,xy}&=&-\frac{2t_s^2}{E_v}\mathbb{I}_{4 \times 4},~
\big[\bs{H}_{ct}^z\big]_{zx,yz}=-\frac{2t_s^2}{E_v}\mathbb{I}_{4 \times 4},~
\big[\bs{H}_{ct}^z\big]_{zx,zx}=-\frac{2t_s^2}{E_v}\mathbb{I}_{4 \times 4}.
\eea
		
Now we consider terms associated with Eq.~(\ref{eq:ct2_2}). In each subspace, the contribution is given as
\bea
\nonumber
\big[\bs{H}_{ct}^z\big]_{xy,xy}&=&-\frac{2t_s^2}{E_v}\mathbb{I}_{4 \times 4},~
\big[\bs{H}_{ct}^z\big]_{xy,yz}=-\frac{4t_s^2}{E_v}\mathbb{I}_{4 \times 4},~
\big[\bs{H}_{ct}^z\big]_{xy,zx}=-\frac{4t_s^2}{E_v}\mathbb{I}_{4 \times 4},\\
\nonumber
\big[\bs{H}_{ct}^z\big]_{yz,xy}&=&-\frac{t_s^2}{E_v}\mathbb{I}_{4 \times 4},~
\big[\bs{H}_{ct}^z\big]_{yz,yz}=-\frac{2t_s^2}{E_v}\mathbb{I}_{4 \times 4},~
\big[\bs{H}_{ct}^z\big]_{yz,zx}=-\frac{2t_s^2}{E_v}\mathbb{I}_{4 \times 4},\\
\big[\bs{H}_{ct}^z\big]_{zx,xy}&=&-\frac{2t_s^2}{E_v}\mathbb{I}_{4 \times 4},~
\big[\bs{H}_{ct}^z\big]_{zx,yz}=-\frac{4t_s^2}{E_v}\mathbb{I}_{4 \times 4},~
\big[\bs{H}_{ct}^z\big]_{zx,zx}=-\frac{4t_s^2}{E_v}\mathbb{I}_{4 \times 4}.
\eea
The contribution from Eq.~(\ref{eq:ct2_3}) in each subspace is given as
\bea
\nonumber
\big[\bs{H}_{ct}^z\big]_{xy,xy}&=&-\frac{2t_s^2}{E_v}\mathbb{I}_{4 \times 4},~
\big[\bs{H}_{ct}^z\big]_{xy,yz}=-\frac{2t_s^2}{E_v}\mathbb{I}_{4 \times 4},~
\big[\bs{H}_{ct}^z\big]_{xy,zx}=-\frac{t_s^2}{E_v}\mathbb{I}_{4 \times 4},\\
\nonumber
\big[\bs{H}_{ct}^z\big]_{yz,xy}&=&-\frac{4t_s^2}{E_v}\mathbb{I}_{4 \times 4},~
\big[\bs{H}_{ct}^z\big]_{yz,yz}=-\frac{4t_s^2}{E_v}\mathbb{I}_{4 \times 4},~
\big[\bs{H}_{ct}^z\big]_{yz,zx}=-\frac{2t_s^2}{E_v}\mathbb{I}_{4 \times 4},\\
\big[\bs{H}_{ct}^z\big]_{zx,xy}&=&-\frac{4t_s^2}{E_v}\mathbb{I}_{4 \times 4},~
\big[\bs{H}_{ct}^z\big]_{zx,yz}=-\frac{4t_s^2}{E_v}\mathbb{I}_{4 \times 4},~
\big[\bs{H}_{ct}^z\big]_{zx,zx}=-\frac{2t_s^2}{E_v}\mathbb{I}_{4 \times 4}.
\eea
The contribution from Eq.~(\ref{eq:ct2_4}) in each subspace is given as
\bea
\nonumber
\big[\bs{H}_{ct}^z\big]_{xy,xy}&=&-\frac{t_s^2}{E_v}\mathbb{I}_{4 \times 4},~
\big[\bs{H}_{ct}^z\big]_{xy,yz}=-\frac{2t_s^2}{E_v}\mathbb{I}_{4 \times 4},~
\big[\bs{H}_{ct}^z\big]_{xy,zx}=-\frac{2t_s^2}{E_v}\mathbb{I}_{4 \times 4},\\
\nonumber
\big[\bs{H}_{ct}^z\big]_{yz,xy}&=&-\frac{2t_s^2}{E_v}\mathbb{I}_{4 \times 4},~
\big[\bs{H}_{ct}^z\big]_{yz,yz}=-\frac{4t_s^2}{E_v}\mathbb{I}_{4 \times 4},~
\big[\bs{H}_{ct}^z\big]_{yz,zx}=-\frac{4t_s^2}{E_v}\mathbb{I}_{4 \times 4},\\
\big[\bs{H}_{ct}^z\big]_{zx,xy}&=&-\frac{2t_s^2}{E_v}\mathbb{I}_{4 \times 4},~
\big[\bs{H}_{ct}^z\big]_{zx,yz}=-\frac{4t_s^2}{E_v}\mathbb{I}_{4 \times 4},~
\big[\bs{H}_{ct}^z\big]_{zx,zx}=-\frac{4t_s^2}{E_v}\mathbb{I}_{4 \times 4}.
\eea
\begin{figure}[h!]
	\includegraphics[width=0.3\columnwidth]{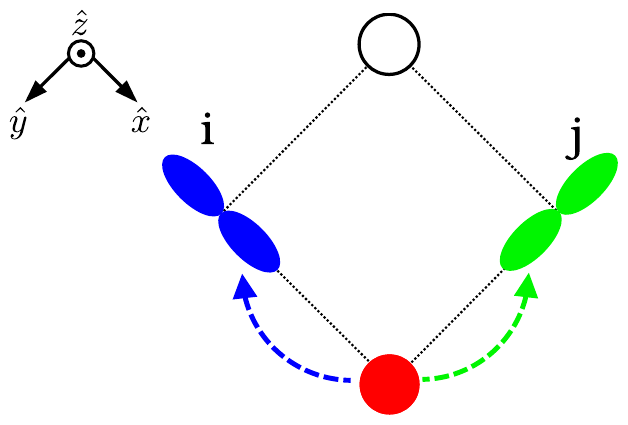}
	\caption{
	}
\label{fig:ct3_2}
\end{figure}
One also needs to take into account symmetric processes $(yz \leftrightarrow zx)$ with another anion as illustrated in Fig.~\ref{fig:ct3_2}. Collecting all possible charge-transfer contributions, we obtain the effective spin-orbital Hamiltonian,
\bea
H_{ct} \!=\! \frac{2t_s^2}{2U + 2\Delta - 9 U_p}  \sum_{\langle ij \rangle \parallel \gamma}
\!\!\Big( \vec{S}_i \! \cdot \! \vec{S}_j - \frac{1}{4}\Big) (n_{i}^{a}n_{j}^{b}\!+\!n_{i}^{b}n_{j}^{a}).
\label{eq:H_ct_s}
\eea
		
\subsection{II. Detail of infinite density matrix renormalization group simulation}
To perform the infinite matrix product state related simulation, we first clarify the ordering of lattice sites, which is implemented by labeling an integer $i$ to each lattice site. We follow a site labeling method for honeycomb lattice on the YC-2$L_y$ cylinders as shown in Fig.~\ref{fig:cylinder}. Note that YC-4 cylinder cannot support the Kekule spin-orbit dimerized phase.
		
\begin{figure}[h]
	\centering
	\includegraphics[scale=1.2]{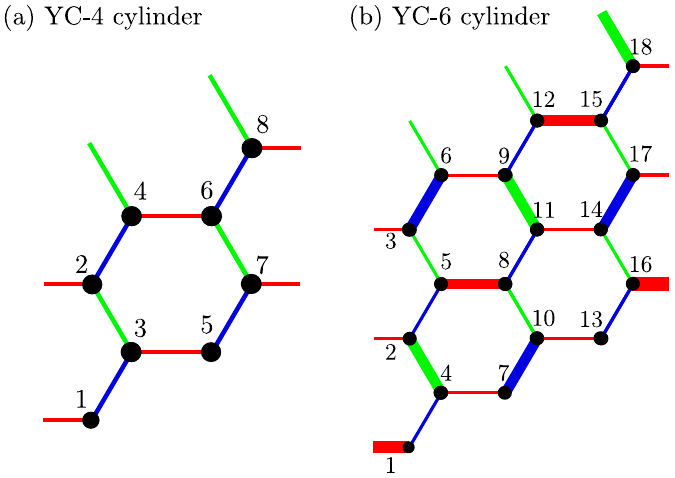}
	\caption{(color online) (a,b) Site-labeling scheme for a honeycomb lattice on a YC-4 or YC-6 cylinder with $x$ direction length $L_x=2$ or 3, respectively. The total number of sites is $2L_xL_y = $ 8 or 18, respectively. We wrap the honeycomb lattice onto a cylinder which is periodic along $y$ direction and infinite along $x$ direction.} 
	\label{fig:cylinder}
\end{figure}
		
To check the convergence of the ground state obtained by iDMRG, we investigate the scaling behavior of the ground state energy per site as a function of bond dimension $\chi$ on the YC-6 cylinder with $L_x=3$. Fig.~\ref{fig:scaling} shows that the per site energy $E_0/J_d$ scales linearly with $1/\chi$ indicating a reliable convergence. 
		
\begin{figure}[h]
	\centering
	\includegraphics[scale=1]{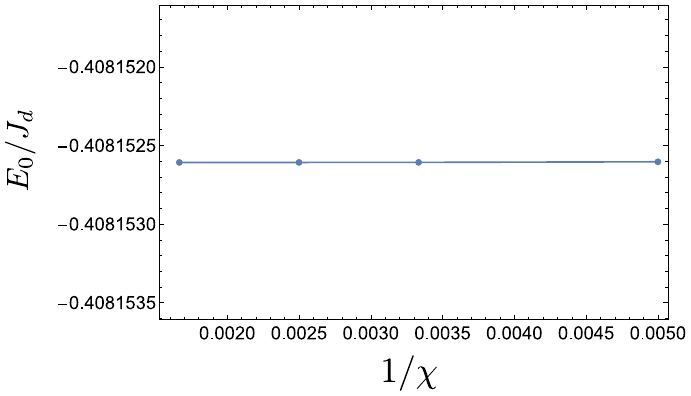}
	\caption{(color online) Per site ground state energy $E_0/J_d$ as a function of bond dimension $\chi$ on a YC-6 cylinder with length $L_x$ = 3.} 
	\label{fig:scaling}
\end{figure}

\subsection{III. Detail of exact diagonalization simulation}
In this section, we give a detail of ED simulation on a cluster of six sites with periodic boundary conditions as shown in Fig.~\ref{fig:ED}(a). The cluster can support the Kekule as well as the columnar spin-orbit dimerized phase and the ground state turns out to be the Kekule phase. In Fig.~\ref{fig:ED}(b), we plot the energy per site $E_0$ as a function of $\tilde J$ for the ground state, Kekule phase, and the first excited state. The result indicates the stabilization of the Kekule phase up to $\tilde J=0.5$.
\begin{figure}[h]
	\centering
	\includegraphics[scale=1.2]{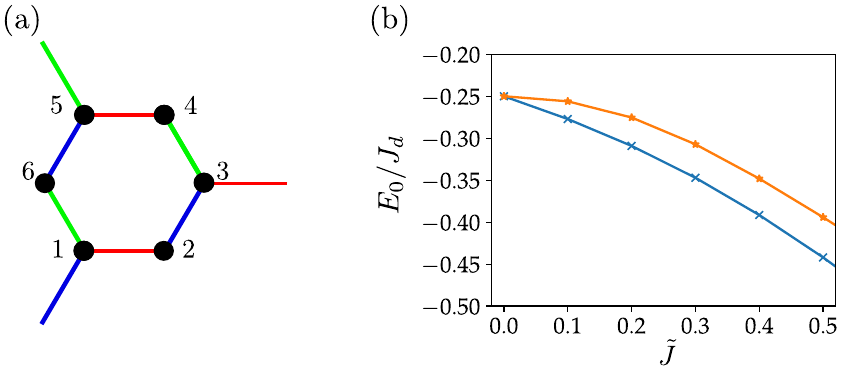}
	\caption{(color online) (a) Six sites cluster with periodic boundary conditions which can support the Kekule phase. (b) Energy per site $E_0$ for the ground state, which is the Kekule phase, and the first excited state.} 
	\label{fig:ED}
\end{figure}
		
\subsection{IV. Gell-Mann matrices}
In this section, we give the explicit expressions of Gell-Mann matrices $\lambda_i$.		
\begin{gather}
	\lambda_1\! =\!
	\!\!\left(\begin{array}{ccc}
		0 & 1 & 0\\
		1& 0 & 0\\
		0 & 0 & 0
	\end{array}\right)\!\!,~
	\lambda_2\! =\!
	\!\!\left(\begin{array}{ccc}
		0 & -i & 0\\
		i & 0 & 0\\
		0 & 0 & 0
	\end{array}\right)\!\!,~
	\lambda_3=\!
	\!\!\left(\begin{array}{ccc}
		1 & 0 & 0\\
		0& -1 & 0\\
		0 & 0 & 0
	\end{array}\right)\!\!,~
	\lambda_4\! =\!
	\!\!\left(\begin{array}{ccc}
		0 &0 & 1\\
		0& 0 & 0\\
		1 & 0 & 0
	\end{array}\right)\!\!,\nonumber \\
	\lambda_5\! =\!
	\!\!\left(\begin{array}{ccc}
		0 & 0& i\\
		0& 0 & 0\\
		-i & 0 & 0
	\end{array}\right)\!\!,~
	\lambda_6=\!
	\!\!\left(\begin{array}{ccc}
		0 & 0 & 0\\
		0& 0 & 1\\
		0 & 1 & 0
	\end{array}\right)\!\!,~
	\lambda_7\! =\!
	\!\!\left(\begin{array}{ccc}
		0 &0 & 0\\
		0& 0 & -i\\
		0 & i & 0
	\end{array}\right)\!\!,~
	\lambda_8\! =\! \frac{1}{\sqrt{3}}
	\!\!\left(\begin{array}{ccc}
		1 & 0& 0\\
		0& 1& 0\\
		0& 0 & -2
	\end{array}\right)\!\!.
\end{gather}

\end{widetext}
	
\end{document}